\documentstyle[epsfig,axodraw,citesort,amsmath]{aipproc}

\newcommand{\pom}{I\!\!P}
\newcommand{\xpom}{x_{\pom}}

\begin{document}
\title{A Very Forward Proton Spectrometer for H1\thanks{Talk presented at LISHEP 2002, Session C: Workshop on Diffractive Physics, February 4-8, 2002, Rio de Janeiro, RJ, Brazil.}}
\author{Pierre Van Mechelen\thanks{
Postdoctoral Fellow of the Fund for Scientific Research - Flanders (Belgium)}}
\address{Physics Department \\
 University of Antwerpen (UIA)  \\
Universiteitsplein 1 - 2610 Antwerpen - Belgium \\
E-mail: Pierre.VanMechelen@ua.ac.be}

%\lefthead{Pierre Van Mechelen}
%\righthead{A Very Forward Proton Spectrometer for H1}
\maketitle

\begin{abstract}
A new, very forward, proton spectrometer with large acceptance will be installed in the proton beam line of the H1 experiment in 2003.  The spectrometer, located 220~m downstream of the interaction point, is based on the Roman Pot technique and consists of two stations in the cold section of the proton beam line.  A brief description of this new device and expected physics results are presented.
\end{abstract}

\section*{Introduction}

%\subsection*{Diffraction Physics at HERA}

The observation during the first running period of HERA (1992-2000) of events with a large rapidity gap in the hadronic final state~\cite{bib:hera_lrgobs}, which are attributed to diffractive dissociation of (virtual) photons, led to a renewed interest in the study of diffraction and its underlying dynamics. In particular, the expectation is that the reconciliation of approaches based on Regge phenomenology on the one hand and on quantum chromodynamics (QCD) on the other hand will lead to a better understanding of QCD in the limit of small scales and/or high parton densities.

Conclusions drawn from measurements of the inclusive cross section~\cite{bib:hera_inclxsec} have been confirmed by studies of the properties of the hadronic final state of diffractive virtual photon dissociation, covering inclusive properties like event shapes and particle spectra~\cite{bib:hera_inclhfs}, semi-inclusive jet and open charm production~\cite{bib:hera_semihfs} and exclusive vector meson production~\cite{bib:hera_exclvm}.  The main results of these analyses are the establishment of a transition from soft to hard diffraction, an interaction dynamics dominated by the presence of gluons and parton densities adhering to DGLAP equations.

Although the experimental results on diffraction from the \mbox{HERA-I} running period are abundant, they suffer from major limitations.  The bulk of diffractive events are selected using a rapidity gap criterion yielding large systematic errors due to the less well known efficiencies of forward detectors and due to theoretical uncertainties on the double dissociation background.  Results are also still statistically limited, especially when requiring the presence of jets, charmed particles or heavy vector mesons.  Finally, for the majority of the detected events, the scattered proton momentum is not measured, thereby limiting the study of the distributions of momentum transfer $|t|$ and azimuthal scattering angle $\phi$.

During the 2000/2001 shutdown, both the HERA accelerator~\cite{bib:hera_upgrade} and the H1 detector have been upgraded, yielding an increase in luminosity by a factor of 5 and improving triggering capabilities for jets and heavy quarks. As explained below, the high luminosity regime also necessitates an efficient trigger for diffractive interactions.  A Very Forward Proton Spectrometer (VFPS) has therefore been proposed~\cite{bib:vfps_proposal} which will provide a clean selection of diffractive events by tagging the scattered proton.  In addition, it will become possible to measure the proton momentum and thus extract the full set of kinematic variables $\xpom$, $|t|$ and $\phi$.

\subsection*{The H1 Forward Proton Spectrometer}

The H1 Collaboration already operated Roman Pot detectors during the \mbox{HERA-I} running period.  These are, however, located much closer to the central H1 detector and have limited acceptance for diffractively scattered protons (i.e.\@ with $\xpom \sim 0.01$). 

Two horizontal stations are positioned at 60 and 80~m from the interaction point and exploit large scattering angles to separate diffractive protons from the nominal beam.  They have an acceptance, integrated over $|t|$, of only a few percent over a wide range of $\xpom$ (see also Tab.\@~\ref{tab:acceptance}). Two vertical stations are located at 81 and 90~m from the interaction point. Here, a vertical dipole magnet is used to separate protons with large energy losses from the nominal beam.  The acceptance is high at large $\xpom$ ($> 0.05$) where, in addition to pure pomeron exchange, also reggeon exchange is expected to contribute.

\begin{table}
\caption{Luminosity and number of events used in physics analyses based on data accumulated with the vertical (FPS-V) and horizontal (FPS-H) Roman Pot stations.}
\label{tab:fpsevents}
\begin{tabular}{@{\hspace{5mm}} l @{\hspace{5mm}} l @{\hspace{5mm}} c @{\hspace{5mm}} r @{\hspace{5mm}} r @{\hspace{5mm}}} 
analysis & detector & year      & luminosity            & events \\ \hline
DIS with leading protons & FPS-V & 1995      & $ 1.4 {\rm\ pb}^{-1}$ & 1661 \\
photoproduction with leading protons & FPS-V & 1996      & $ 3.3 {\rm\ pb}^{-1}$ & 23072 \\
diffractive DIS & FPS-H & 1999/2000 & $28.8 {\rm\ pb}^{-1}$ & 3100 \\
\hline
\end{tabular}
\end{table}

This Forward Proton Spectrometer (FPS)~\cite{bib:fps_detector} has nevertheless yielded interesting physics results~\cite{bib:h1_fps}, albeit based on a limited number of events (see Tab.\@~\ref{tab:fpsevents}).  Even when scaling the number of events to the total expected luminosity of the HERA-II running period for which the FPS will be operational ($\sim 350 {\rm\ pb}^{-1}$), a high statistics analysis of diffractive deep-inelastic scattering  is not possible with the present detector set-up. The FPS will continue to be operated during the HERA-II running period.

\subsection*{Triggering Diffractive DIS at HERA-II}

The bulk of diffractive DIS data accumulated during the first running period of the HERA accelerator was obtained using an inclusive trigger strategy. Like for inclusive DIS, the trigger for diffractive DIS events was based on the presence of a substantial energy deposit in the backward calorimeter (SPACAL).  The selection of diffractive events was applied off-line using a rapidity gap criterion.

With the commissioning of HERA-II the luminosity is expected to increase by a factor 5 while the proton beam current will increase by a factor 2. Inclusive triggers will be disabled or heavily pre-scaled to keep the total trigger rate acceptable.  In order to save diffractive events, an additional trigger requirement  beyond the inclusive SPACAL trigger mentioned above is therefore necessary.  Moving the off-line rapidity gap selection to the trigger level is unfeasible because of large uncertainties in the efficiency calculation of such a trigger and unstable background conditions.  Moreover, while the rate reduction would be considerable for diffractive DIS events, estimates show that the rate reduction from a rapidity-gap trigger for beam-gas and photoproduction background events is marginal~\cite{bib:newman_thesis}.

The best solution is a trigger based on a scattered proton tag.  It has been estimated that, despite the large distance (220~m) of the VFPS from the H1 detector, it is still possible to use VFPS information at the first trigger level.  Such a VFPS trigger, used in combination with the inclusive SPACAL trigger, would decrease the rate of background and physics triggered events by approximately a factor 175, yielding an acceptable trigger rate of 1~Hz.

Note that the upgraded central H1 detector already allows to directly trigger on open charm, jets or heavy vector mesons.  But also in this case the VFPS can help to keep trigger conditions as loose as possible and to lower the rate, especially for dijet production at low $p_T$.

\section*{The Very Forward Proton Spectrometer}

\begin{figure}
\centerline{
\begin{picture}(380,538)
\Text(190,269)[]{\psfig{file=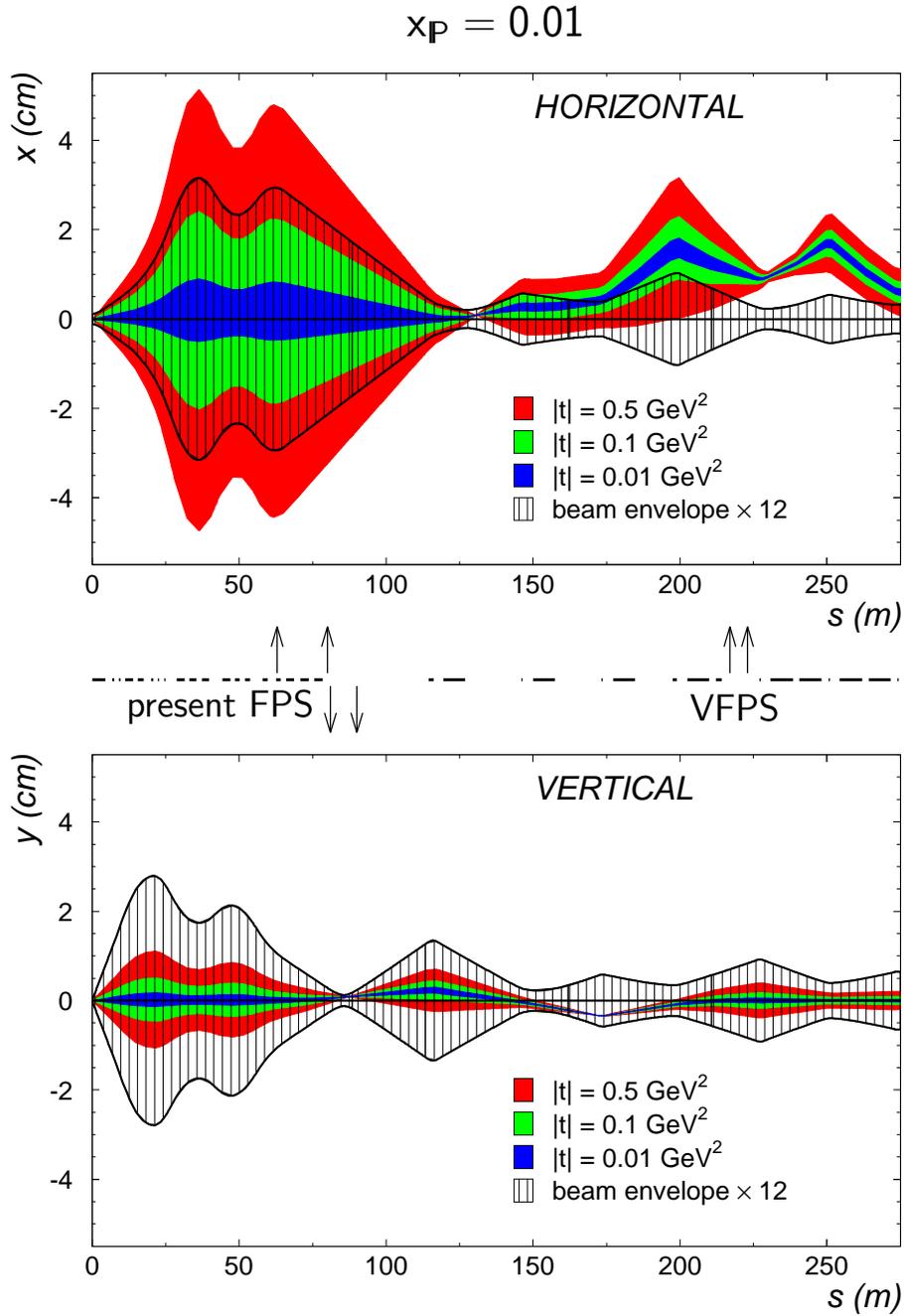,width=380pt}}
\Text(85,260)[]{\large\sf present FPS}
\Text(281,260)[]{\large\sf VFPS}
\Text(190,520)[]{\Large${\mathsf{x_{I\!P} = 0.01}}$}
\end{picture}}
\caption{Horizontal (top) and vertical (bottom) proton trajectories for $\xpom = 0.01$ and different $|t|$ values as a function of the position $s$ along the beam line.  The hatched area represents the beam envelope multiplied by 12.  The dots between the two figures indicate the position of magnet elements; white spaces are drift sections.  The position of the present FPS and the future VFPS are also indicated.}
\label{fig:optics}
\end{figure}

The search for an optimal location for the VFPS was guided by the need for a large acceptance for protons with an energy loss of around 1\%.  Figure~\ref{fig:optics} shows the trajectories of protons with $\xpom = 0.01$ and different values of $|t|$ together with the beam envelope multiplied by 12 and the location of magnet elements.  The ``diffractive beam'' leaves the nominal beam envelope horizontally at a drift segment around 220~m.  The Roman Pots will therefore approach the beam horizontally (down to a distance of 5 mm) from inside the HERA ring, exploiting the spectrometer effect of the HERA bend.

The location at 220 m downstream of the H1 interaction point is in the cold (super-conducting) section of the HERA ring.  A horizontal bypass therefore needs to be built for the helium and superconductor lines, so that the Roman Pot detectors can operate at room temperature and freely approach the proton beam.

\subsection*{Detectors}

The VFPS detectors are similar to the ones used in the FPS.  Two VFPS Roman Pot stations, approximately 4~m apart, will be equipped with two scintillating fibre detectors each, sandwiched between trigger planes.  One such fibre detector measures two space coordinates perpendicular to the beam.  The two-fold structure within one Roman Pot stations allows to suppress background hits by selecting forward going tracks.

\begin{figure}
\centerline{\psfig{file=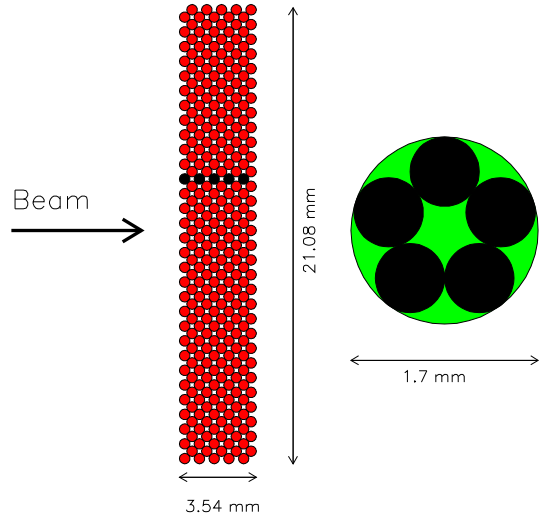,height=80mm}}
\caption{A single plane with staggered scintillating fibres (left) and the arrangement of 5 fibres to be connected to one light guide (right).}
\label{fig:detector}
\end{figure}

The scintillating fibres are staggered as shown in Fig.\@~\ref{fig:detector}.  A diffractive proton (parallel to the beam axis) hits five consecutive fibres which are all connected to a position-sensitive photo-multiplier through the same light guide, yielding on average 8.2 photo-electrons and a 99.4~\% detection efficiency.  The staggered fibre properties are listed in Tab.\@~\ref{tab:fibres}.  The spatial resolution in real beam conditions has been estimated experimentally by a test run using prototype detectors installed in the FPS~\cite{bib:fps_testrun}.

\begin{table}
\caption{Staggered fibre properties. The theoretical resolution is calculated based on the listed properties; the experimental resolution was determined in a test run.}
\label{tab:fibres}
\begin{tabular}{@{\hspace{5mm}} l @{\hspace{30mm}} r @{\hspace{5mm}}} 
diameter                  & $480 {\rm\ \mu m}$ \\
pitch                     & $340 {\rm\ \mu m}$ \\
cladding                  & $30  {\rm\ \mu m}$ \\
theoretical resolution    & $63  {\rm\ \mu m}$ \\
experimental resolution   & $94  {\rm\ \mu m}$ \\
\hline
\end{tabular}
\end{table}

\subsection*{Simulation Studies}

In order to determine the optimal detector location as well as the acceptances and resolutions, proton trajectories have been simulated using a linear beam transport method for the description of the HERA beam optics.  In such an approach, the position ($x$), slope ($x'$) and energy deviation ($\xi = \tfrac{(E-E_{beam})}{E_{beam}}$) of a proton at an arbitrary position along the beam line is related to the position ($x_0$), slope ($x'_0$) and energy deviation ($\xi_0$) at the interaction point and the kinematic variables ($\theta_x$ and $\xpom$) of the diffractive interaction through a set of matrix equations:
\begin{equation}
\left(\begin{matrix}
x \cr x' \cr \xi
\end{matrix}\right)
= 
\left(\begin{matrix}
T_x^{11} & T_x^{12} & D_x \cr
T_x^{21} & T_x^{22} & D'_x \cr
0        & 0        & 1
\end{matrix}\right)
\cdot
\left[
\left(\begin{matrix}
x_0 \cr x'_0 \cr \xi_0
\end{matrix}\right)
+
\left(\begin{matrix}
0 \cr \theta_x \cr - \xpom
\end{matrix}\right)
\right].
\end{equation}
Here $T_x^{11}$, $T_x^{12}$, $T_x^{21}$, $T_x^{22}$ and $D_x$, $D'_x$ are matrix elements depending on the beam optics which describe, to first order, the effect of varying the initial proton position, slope or energy deviation.  A similar equation holds for the vertical coordinate $y$; the motion of the proton in the horizontal and vertical planes are therefore independent.

In principal, this approach is valid for a beam-line consisting of only dipole and quadrupole magnets since sextupole and higher order magnets will introduce non-linearities and correlations between the horizontal and vertical planes.

In addition to the above calculations, non-linear corrections have been applied and found to be important.  These include non-linear effects in the dependence of the energy deviation, the effect of sextupole magnets and the proper description of magnet offsets and tilts w.r.t.\@ to the nominal proton trajectory.

\subsubsection*{Acceptance}

A detailed simulation of the beam pipe shape and apertures has been included in the calculation of the VFPS acceptances. 
%It was found to be necessary (and possible) to introduce a so-called local {\em orbit bump} to steer the beam away from critical acceptance losses.
The acceptances are calculated for an approach of the detectors to the nominal beam of 12 times the beam width and with an additional safety margin of 3~mm for the so-called coasting beam.  This coasting beam consists of particles that have leaked out of the RF buckets and that are therefore no longer accelerated.  They travel around the proton ring until they hit one of the collimators.  The coasting beam can attain the 1~mA level and can potentially cause a substantial background in the VFPS, although these particles are mostly out of time.  Calculations by the HERA machine group have shown that a safety margin of 3~mm would be enough to stay clear of the coasting beam.  It is however still possible that this margin can be lowered in real beam conditions.

%Table~\ref{tab:acceptance} and Fig.\@~\ref{fig:acceptance} show the difference in acceptance between the VFPS only and the combined FPS+VFPS detectors as a function of the variables $|t|$ and $\xpom$.

\begin{figure} 
\vspace{10mm}
\centerline{
\psfig{file=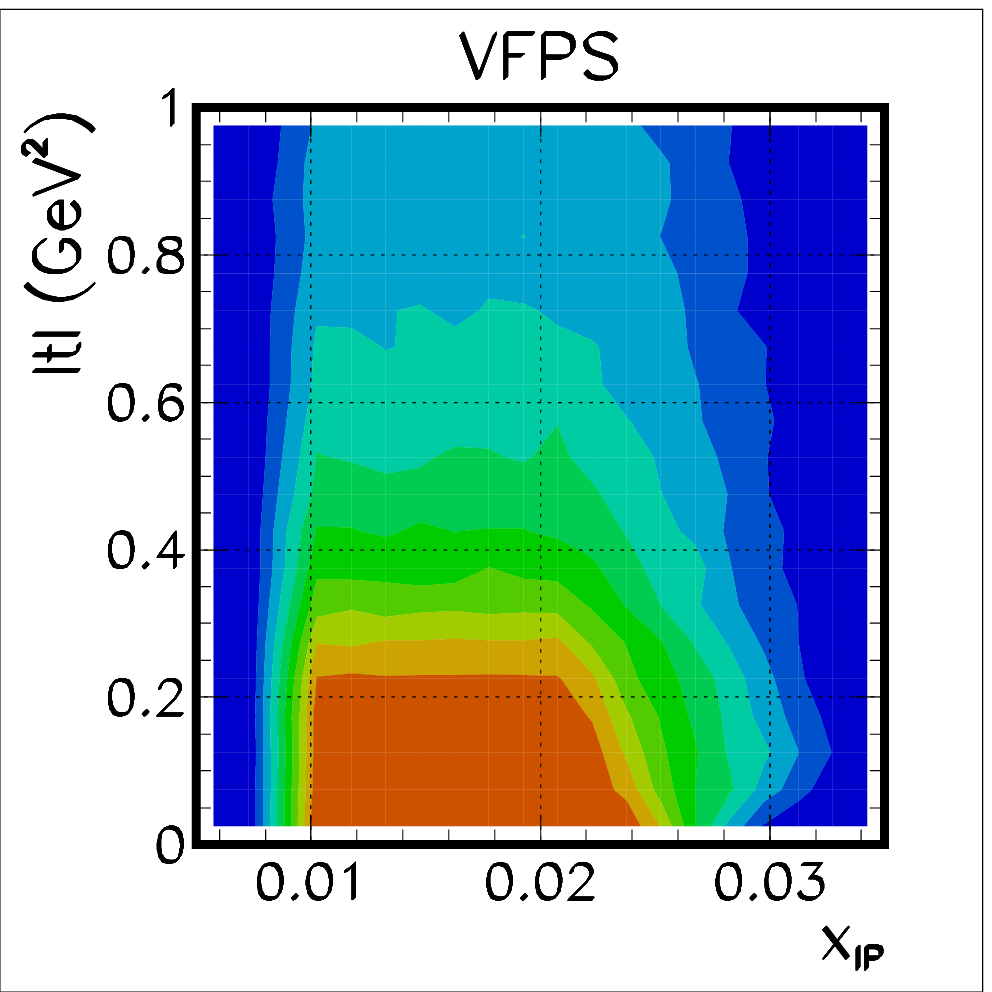,width=0.45\textwidth}\hfil\psfig{file=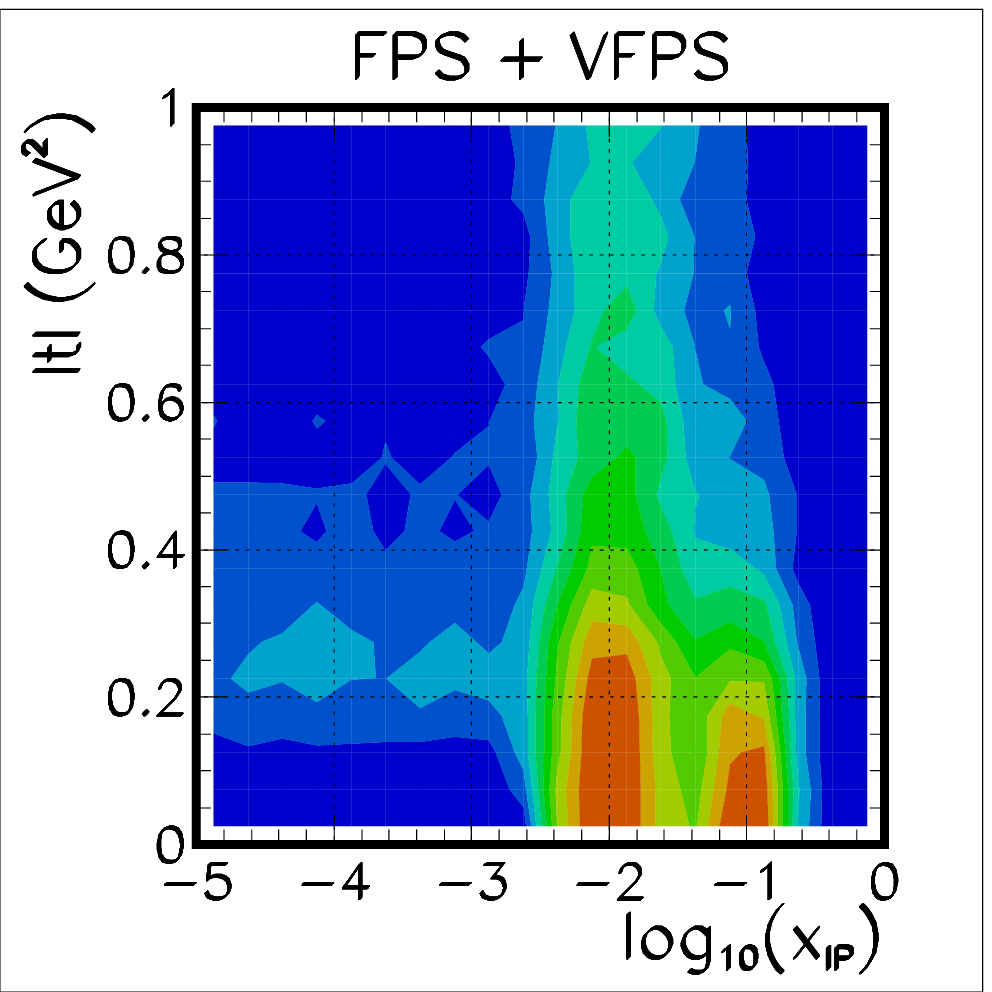,width=0.45\textwidth}}
\vspace{5mm}
\caption{Acceptances as a function of $|t|$ and $\xpom$ for the VFPS (left) and for the vertical and horizontal FPS and VFPS combined (right).  The colour coding is in steps of 10~\%, blue representing 0 to 10~\% and red 90 to 100~\%.  The acceptance in the left figure has been calculated as described in the text.  For the right figure, the coasting beam safety margin has been dropped and the detectors are allowed to approach the proton beam down to 8 times the beam width (this is how the FPS has been operated in the past).}
\label{fig:acceptance}
\end{figure}

The acceptance range of the VFPS compared to the acceptance of the horizontal and vertical FPS are listed in Tab.\@~\ref{tab:acceptance} and shown in Fig.\@~\ref{fig:acceptance}.

\begin{table}
\caption{Acceptance ranges in $|t|$ and $\xpom$ and average local acceptances within these ranges for the horizontal and vertical FPS and for the VFPS.}
\label{tab:acceptance}
\begin{tabular}{@{\hspace{5mm}} l @{\hspace{5mm}} l @{\hspace{5mm}} l @{\hspace{5mm}} l @{\hspace{5mm}}} 
 & FPS-H & FPS-V & VFPS \\ \hline
$|t|$-range & 0.2 $\rightarrow$ 0.4 & 0 $\rightarrow$ 0.15 & 0 $\rightarrow$ 0.25 \\ 
$\xpom$-range & $10^{-5} \rightarrow 10^{-2}$ & 0.05 $\rightarrow$ 0.15 & 0.01 $\rightarrow$ 0.02 \\
local acceptance & $<$ 30 \% & $\sim$ 100 \% & $\sim$ 100 \% 
\end{tabular}
\end{table}

\subsubsection*{Resolution}

\begin{figure}
\centerline{\psfig{file=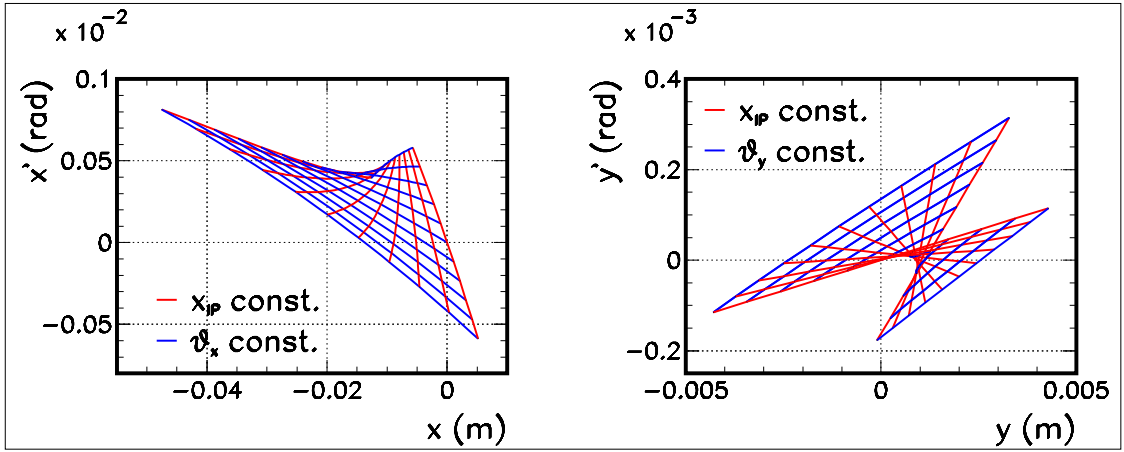,width=\textwidth}}
\vspace{5mm}
\caption{Mapping of the kinematic variables describing the diffractive interaction onto the coordinates planes $x$-$x'$ and $y$-$y'$ at the location of the VFPS ($-1 {\rm\ mrad} < \theta_x, \theta_y < +1 {\rm\ mrad}$ and $0 < \xpom < 0.03$).}
\label{fig:mapping}
\end{figure}

For the reconstruction of kinematic variables describing the diffractive proton ($\theta_x$, $\theta_y$ and $\xpom$) one has to take into account the mapping of these variables onto the $x$-$x'$ and $y$-$y'$ planes as displayed in Fig.\@~\ref{fig:mapping}.

This is achieved by minimizing a $\chi^2$ defined as:
\begin{equation}
\chi_{reco}^2 = \left( x_i - x_i(\theta_x, \theta_y, \xpom)\right) \cdot c_{ij}^{-1} \cdot \left( x_j - x_j(\theta_x, \theta_y, \xpom)\right),
\end{equation}
where $x_i$ are the coordinates measured by the VFPS detectors, $x_i(\theta_x, \theta_y, \xpom)$  are the coordinates calculated as a function of the kinematic variables using the beam optics and $c_{ij}$ is a covariance matrix containing the beam covariance and the fibre detector resolution.

\begin{figure}
\centerline{\psfig{file=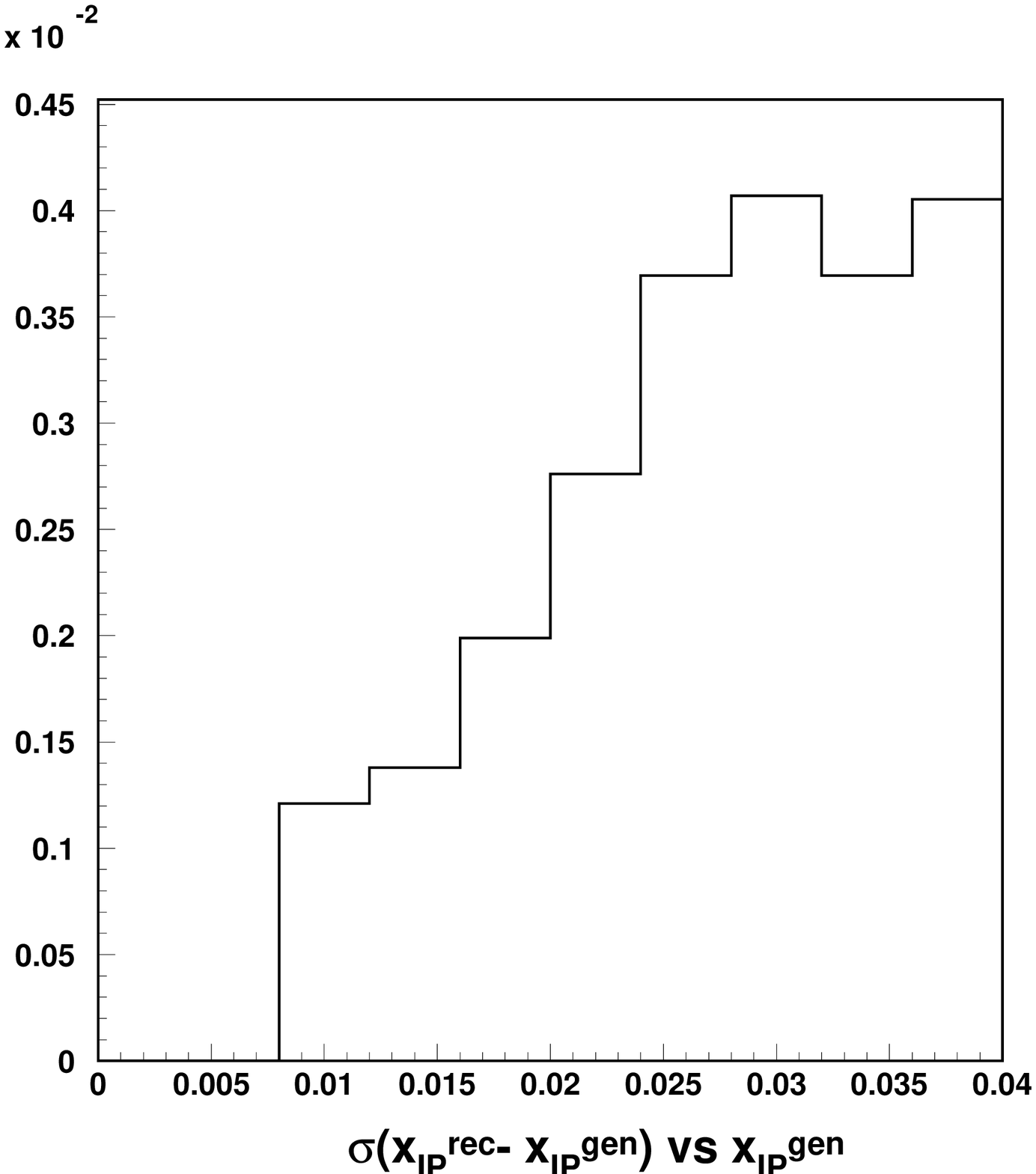,width=0.333\textwidth,height=60mm}
            \psfig{file=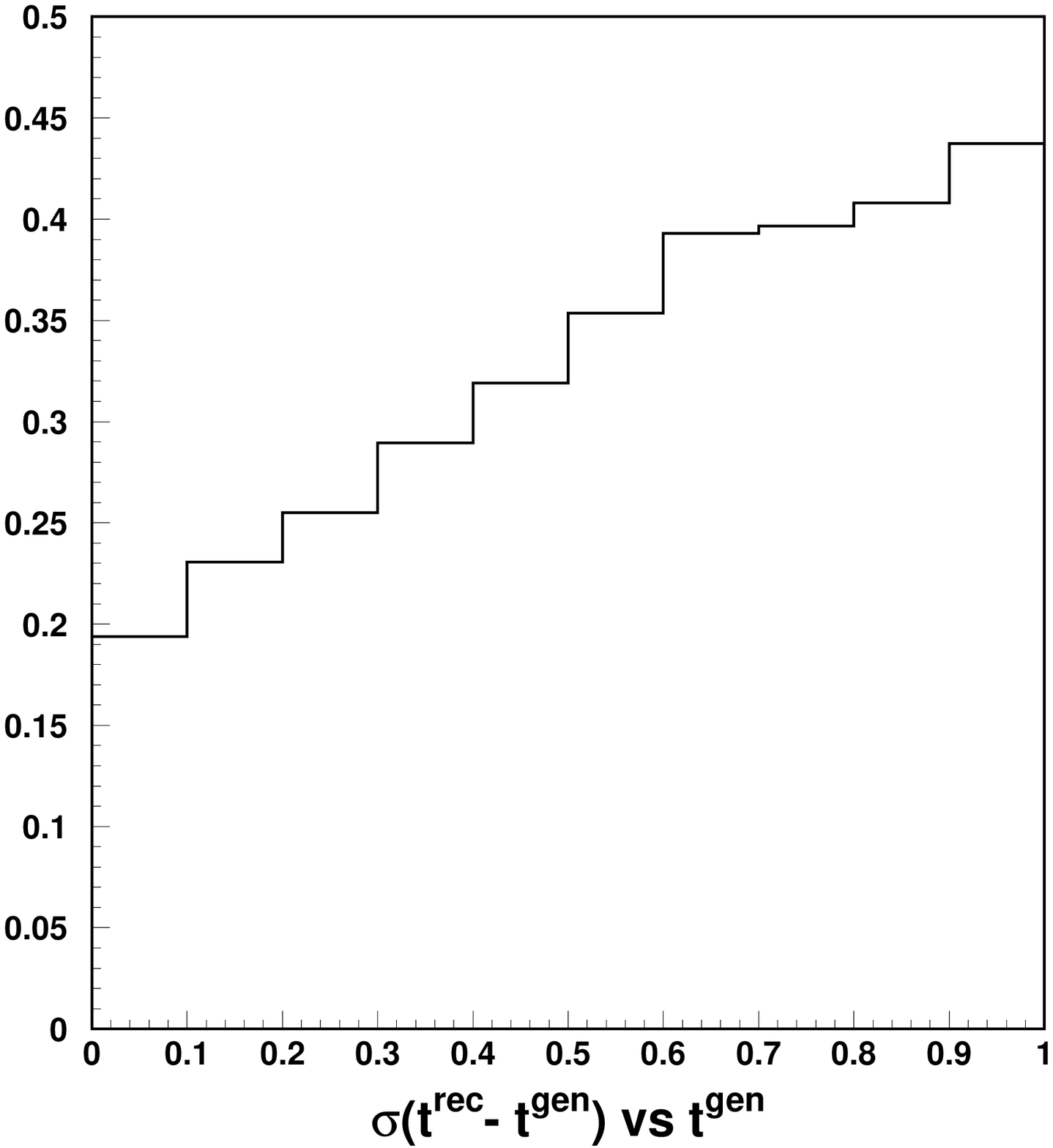,width=0.333\textwidth,height=60mm}
            \psfig{file=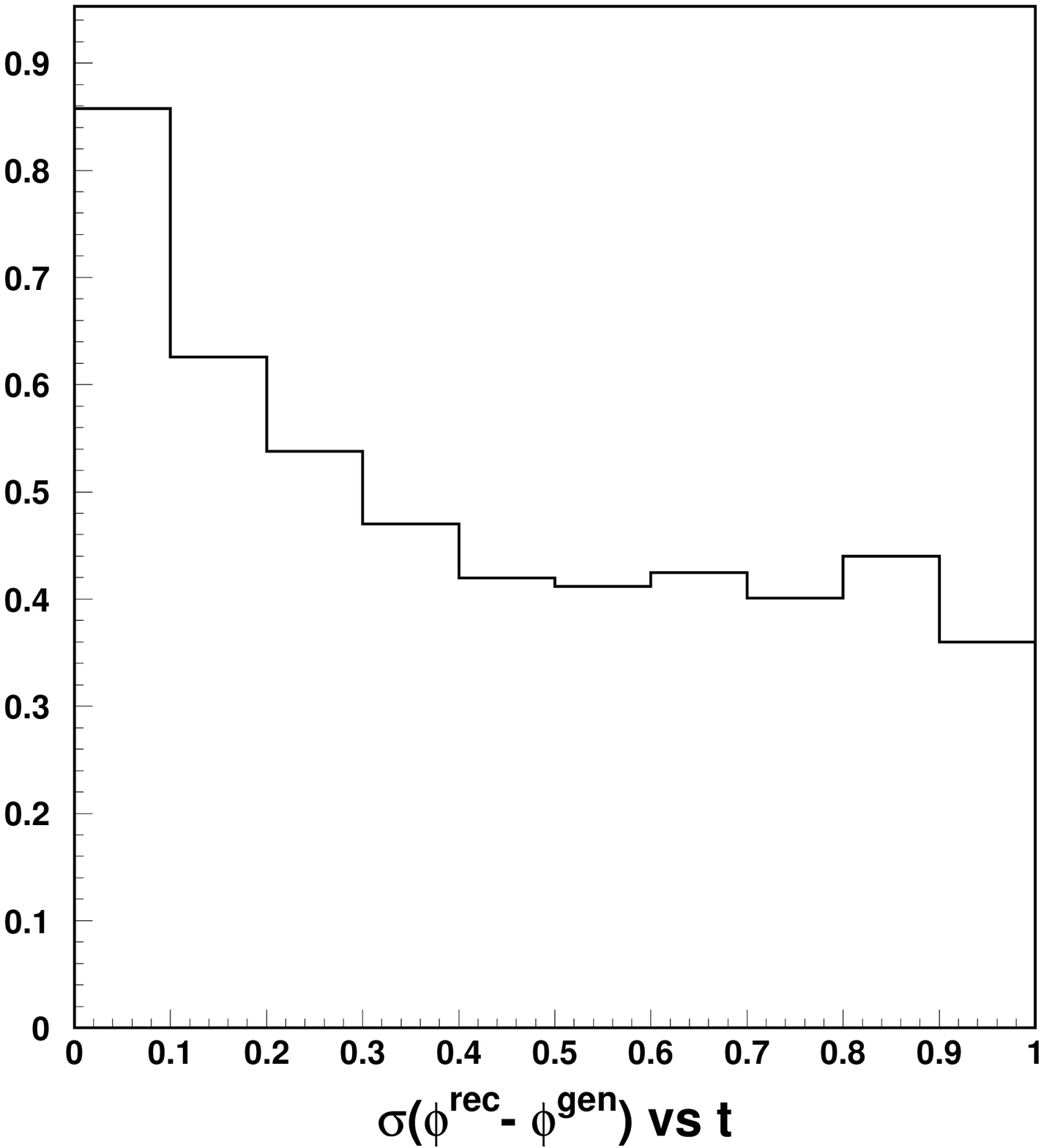, width=0.333\textwidth,height=60mm}}
\caption{Resolutions (root mean squared width of the reconstructed minus generated values) of the diffractive kinematic variables $\xpom$ (left), $|t|$ (middle) and $\phi$ as a function of $\xpom$, $|t|$ and again $|t|$, respectively.}
\label{fig:resolutions}
\end{figure}

Typical resolutions of the reconstructed kinematic variables $\xpom$, $|t|$ and $\phi$ are shown in Fig.\@~\ref{fig:resolutions}.  In general the resolutions are dominated by the beam covariance, although there is still some sensitivity to the fibre detector resolution.  The $\xpom$ resolution is competitive with the reconstruction of $\xpom$ by the central H1 detector.  Furthermore, it will be possible to measure approximately 4 bins in $|t|$ and 15 bins in $\phi$ for $|t| > 0.2 {\rm\ GeV}^2$.

\subsubsection*{Alignment}

The alignment concerns the relative positioning of the VFPS detectors w.r.t.\@ the nominal proton beam.  The distance between the detectors and the beam varies not only due to the Roman Pot movement, which is accurately controlled ($< 5 {\rm\ \mu m}$), but also because the beam position itself may vary between luminosity runs or even within one run.  
In order to exclude any systematic bias due to the detector position, a time-dependent calibration procedure is necessary.  One possible method is to exploit the forward kinematic peak in $\theta_x$ and $\theta_y$ together with the $\xpom$ measurement from the central H1 detector.  In such a procedure another $\chi^2$ minimization is performed with:
\begin{equation}
\chi_{cali}^2 = \frac{\theta_x^2}{\sigma_{\theta_x}^2} + 
                \frac{\theta_y^2}{\sigma_{\theta_y}^2} +
                \frac{\left(\xpom - \xpom^{H1}\right)^2}
                     {\sigma_{\left(\xpom - \xpom^{H1}\right)}^2}.
\end{equation}

\begin{figure}
\begin{minipage}[t][80mm][c]{0.35\textwidth}
\psfig{file=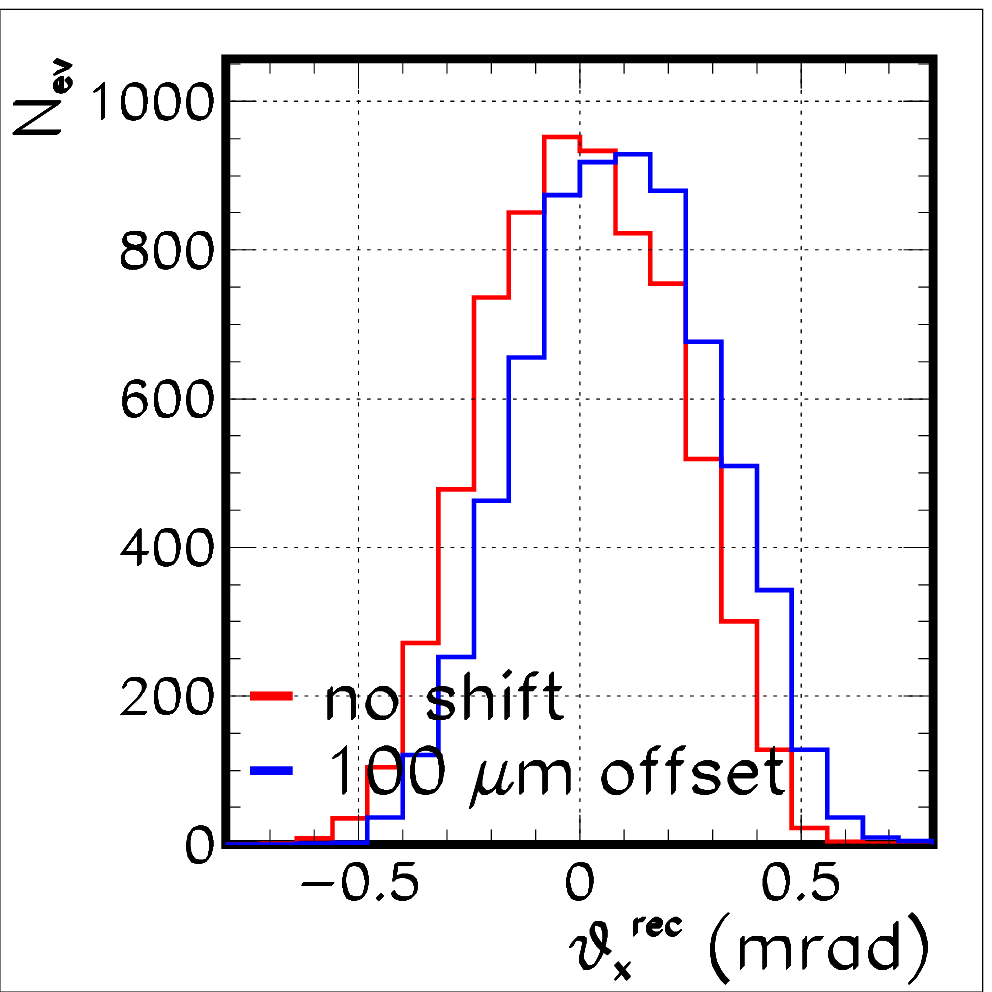,width=\textwidth}
\end{minipage}
\hfil
\begin{minipage}[t][80mm][c]{0.60\textwidth}
\psfig{file=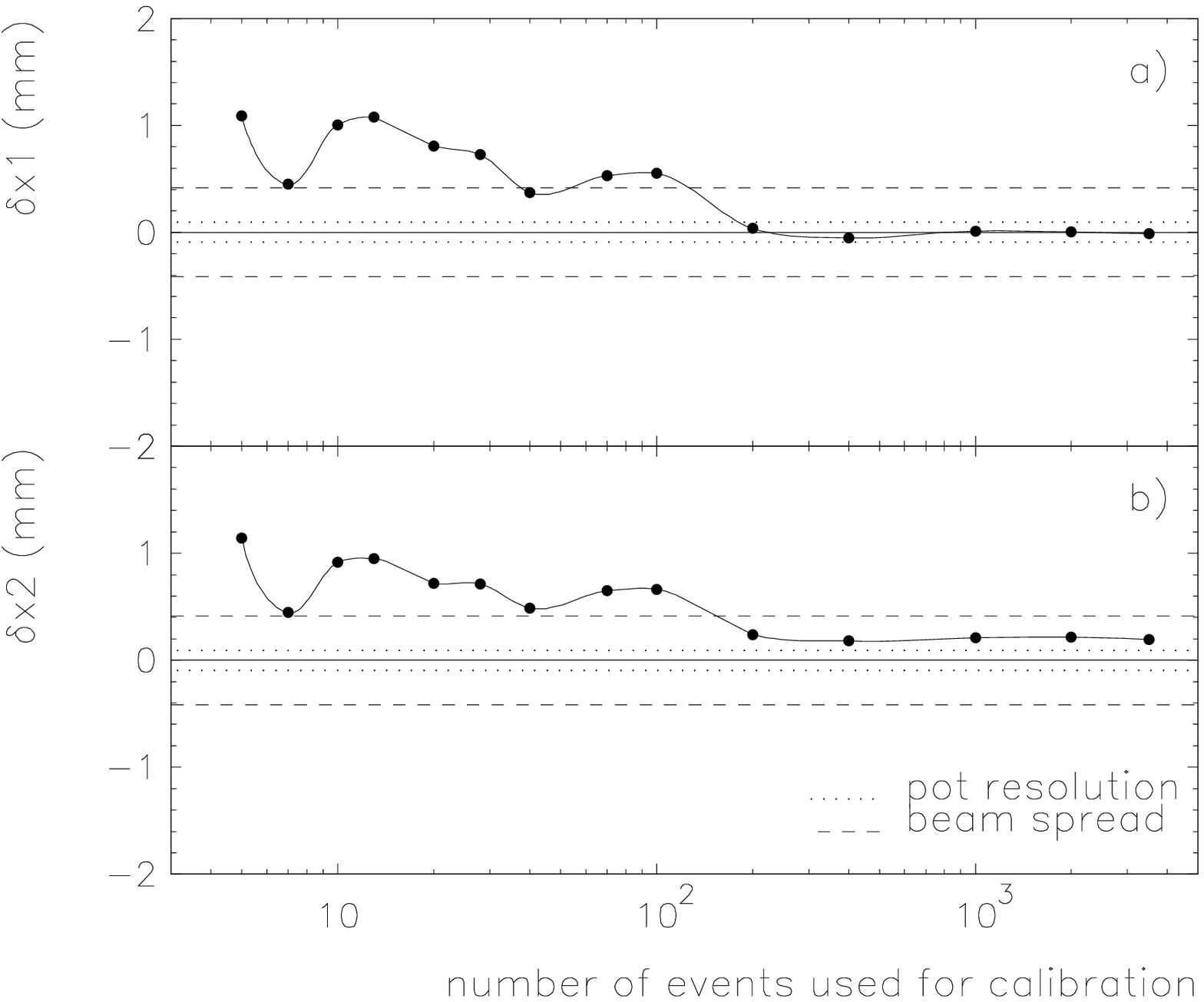,width=\textwidth}
\end{minipage}
\caption{(left) Reconstructed $\theta_x$ distribution for no shift and for a $100 {\rm\ \mu m}$ horizontal offset of one of the Roman Pots. 
(right) Calibrated detector offsets as a function of the number of events used in the algorithm. The second Roman Pot station was displaced horizontally by $100 {\rm\ \mu m}$.  Also shown are the limiting uncertainties due to the intrinsic detector resolutions and beam divergence at the location of the VFPS.}
\label{fig:calibration}
\end{figure}

The sensitivity of the forward kinematic peak in $\theta_x$ to a $100 {\rm\ \mu m}$ offset is shown in Fig.\@~\ref{fig:calibration} (left).  Figure~\ref{fig:calibration} (right) shows the calibrated offsets as a function of the number of events used in the calibration procedure.  From this one can conclude that an alignment precision of $100 {\rm\ \mu m}$ is feasible.

As an additional cross-check, alternative fits can be performed relying on e.g.\@ the kinematic variables calculated from fully reconstructed elastic $\rho$ meson production events.

\subsection*{Expected Results}

A total integrated luminosity of $350 {\rm\ pb}^{-1}$, corresponding to three years of \mbox{HERA-II} running with a 50 \% VFPS operation efficiency, is assumed throughout the following discussion on expected physics results

\subsubsection*{Inclusive Diffraction}

The VFPS will provide an excellent test of hard scattering factorization (i.e.\@ the factorization of the cross section into DGLAP-governed parton density functions and partonic cross sections)~\cite{bib:collins_diff_qcdfact}.  Even taking into account the rather narrow acceptance window in $\xpom$, it will be possible to make an accurate measurement of the structure function $F_2^D(\beta,Q^2)$ at fixed $\xpom$ in bins of, or integrated over, $|t|$.  By measuring $\alpha_{\pom}$ as a function of $Q^2$ it will also be possible to test Regge factorization (i.e.\@ the factorization of the cross section into a factor depending only on $\xpom$ and $|t|$ and another factor depending only on $Q^2$ and $\beta$)~\cite{bib:reggefact}.

\begin{table}
\caption{Estimated event yields available for the measurement of the diffractive structure function $F_2^D(\beta,Q^2)$.  }
\label{tab:eventyields}
\begin{tabular}{@{\hspace{5mm}} l @{\hspace{5mm}} r @{\hspace{5mm}} r @{\hspace{5mm}}}
event sample                   & coasting beam & no coasting beam \\ \hline
acceptance $> 80 \%$           & 1,100,000     & 390,000          \\
$0.0 < |t| < 0.2 {\rm\ GeV}^2$ & 1,800,000     & 810,000          \\
$0.2 < |t| < 0.4 {\rm\ GeV}^2$ &   330,000     & 160,000          \\
$0.4 < |t| < 0.6 {\rm\ GeV}^2$ &    47,000     &  23,000          \\
$0.6 < |t| < 0.8 {\rm\ GeV}^2$ &     6,000     &   3,000          \\
\end{tabular}
\end{table}

\begin{figure}
\centerline{\psfig{file=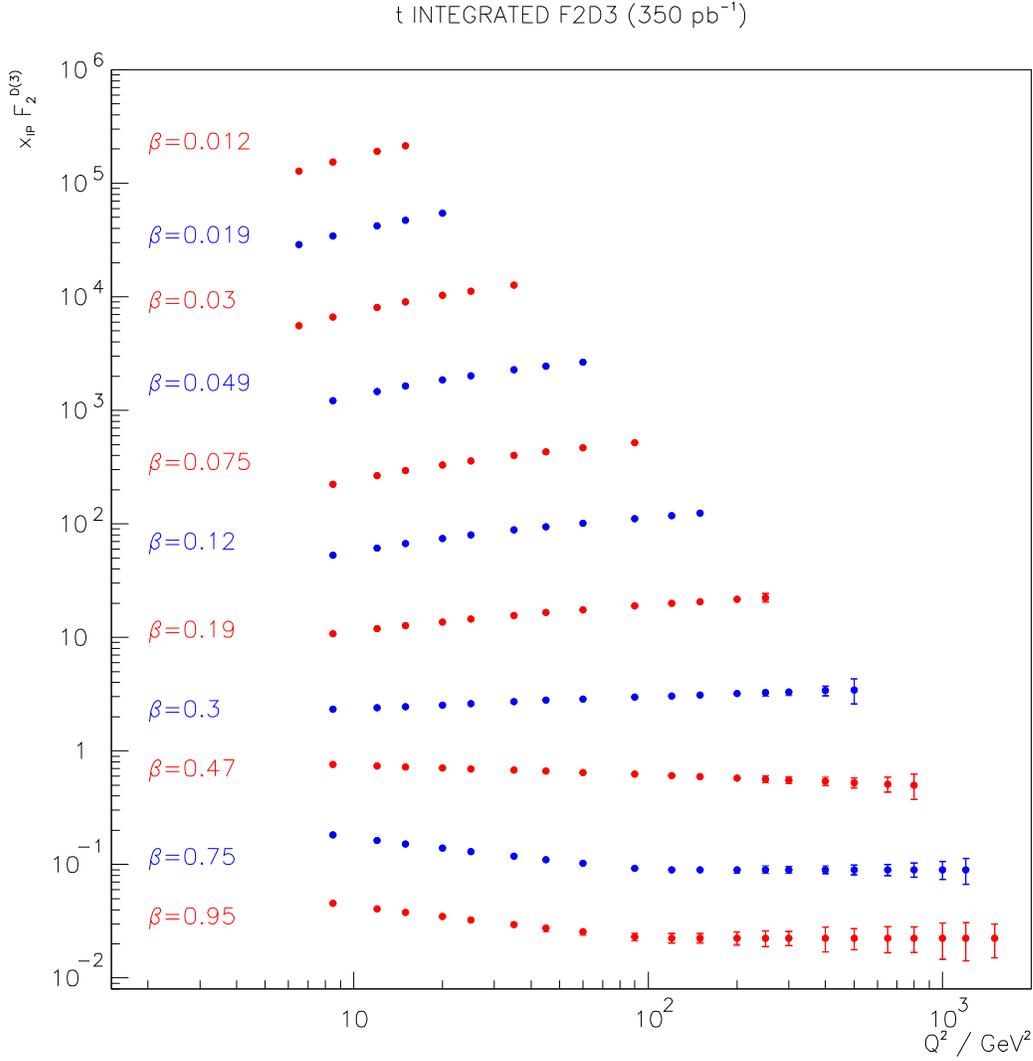,width=\textwidth}}
\caption{Expected precision for $350 {\rm\ pb}^{-1}$ on $F_2^{D(3)}(\beta,Q^2,\xpom)$ at $\xpom = 0.017$, integrated over $|t| < 0.8 {\rm\ GeV}^2$, measured in the region $0.011 < \xpom < 0.024$, where the VFPS acceptance is large, even if the detector has to be $\sim 3.5 {\rm\ mm}$ from the beam due to the coasting beam.  The data points at different $\beta$ are scaled by arbitrary factors for visibility.}
\label{fig:f2d}
\end{figure}

The estimated event yields are listed in Tab.\@~\ref{tab:eventyields} and imply very small statistical errors. Because the scattered proton is directly detected, no systematic errors due to double dissociation background have to be taken into account. Moreover, the uncorrelated systematic errors are expected to approach the level obtained in the measurement of the inclusive structure function $F_2$ (a few \%).  This is because systematic errors associated to the VFPS are common to all data points and thus will result in a global normalization uncertainty. If the effect of the coasting beam can be minimized, two separate bins in $\xpom$ can be studied.  Note, however, that in the case of semi-inclusive studies of jets or charm, the $p_T$ or mass requirement excludes the lower $\xpom$ bin, regardless of the coasting beam situation.  A preview of expected results on $F_2^D$ is shown in Fig.\@~\ref{fig:f2d}.

\subsubsection*{Hadronic Final States}

The study of hadronic final states provides an independent cross-check of the conclusions reached by the analysis of the structure function using the same experimental data (and systematic uncertainties). 

The analysis of open charm production promises to resolve a conflict between results based on 1996/97 data and a prediction by the RAPGAP Monte Carlo~\cite{bib:rapgap} based on a partonic pomeron picture using the parton density function obtained from the measurement of the inclusive structure function.  The 1996/97 measurement was however limited to $46 \pm 10$ events while the analysis of the full \mbox{HERA-I} data sample will only increase this amount by a factor 3 at most.  The \mbox{HERA-II} expectation using the VFPS is 380 events.  About half of all diffractive $D^\ast$ events which can be detected by H1 will have proton tag from the VFPS.

The analysis of diffractive jet electroproduction is still statistically limited in some areas of phase space for 2-jet production and certainly for 3-jet production (only 2500 2-jet and 130 3-jet events were analysed by H1).  The \mbox{HERA-II} expectation using the VFPS is 22900 dijet events.

\begin{figure}
\centerline{\psfig{file=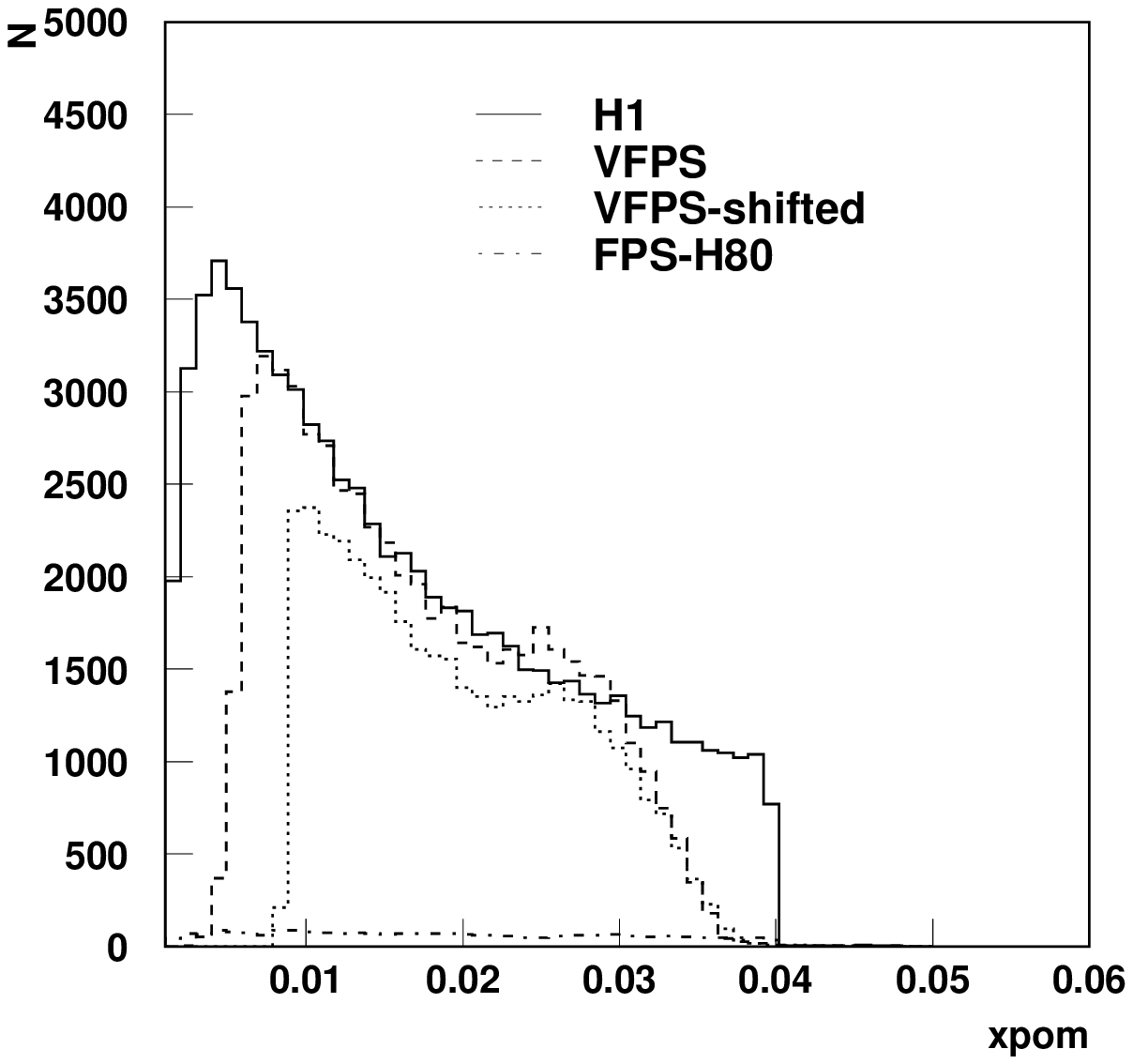,width=0.50\textwidth,height=85mm}
            \psfig{file=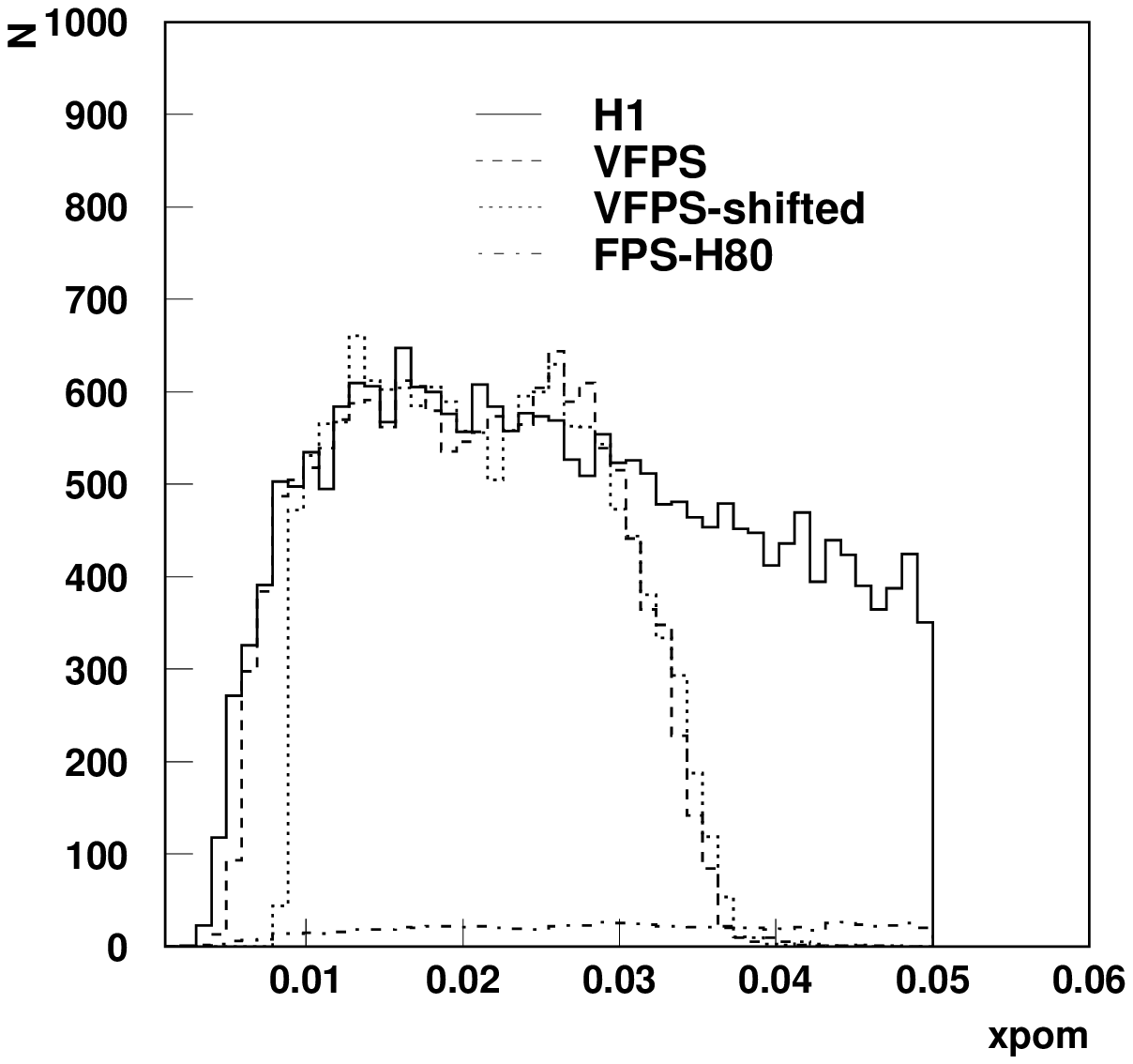,width=0.50\textwidth,height=85mm}}
\caption{$\xpom$ distribution of charm (left) and dijet (right) electroproduction events observable in H1.  The full lines show the distribution for the full H1 central detector acceptance.  The subset of events tagged by the VFPS (without and with an additional shift for the coasting beam) and tagged by the FPS are also shown.  The overall normalization is arbitrary.}
\label{fig:jets_charm}
\end{figure}

Figure~\ref{fig:jets_charm} shows the $\xpom$ distribution of charm and dijet electroproduction events observable in H1.

Another unresolved issue stems from a comparison of HERA and Tevatron results regarding diffractive dijet (photo-) production. Its resolution could provide an insight in the breaking of QCD factorization in hadron-hadron interactions.  Using the H1 Electron Taggers located at 6 and 40~m meters from the interaction point on the outgoing electron side, events with different hadronic centre-of-mass energies ($W$) are selected, allowing to vary the contributions of direct and resolved photons entering the diffractive interaction.  About 20000 (1400) events are expected to be collected at $W = 140\ (275) {\rm\ GeV}$, respectively.

\subsubsection*{Azimuthal Asymmetry}

With a measurement of the azimuthal scattering angle it becomes possible to gain information on the diffractive cross section for longitudinally polarized photons.  The interference between the longitudinal and transverse contributions and between the two different transverse contributions give rise to azimuthal asymmetries~\cite{bib:phiasym}:
\begin{equation}
\frac{{\rm d}\sigma^D}{{\rm d}\Delta\phi} \propto \sigma_T + \sigma_L - 2 \sqrt{\epsilon (1+\epsilon)} \sigma_{LT} \cos \Delta\phi - \epsilon \sigma_{TT} \cos 2\Delta\phi
\end{equation}
where the polarization parameter $\epsilon$ is a function of $y$ only and is very close to unity throughout most of the measurable kinematic range.  $\Delta\phi$ is the angle between the proton and electron scattering planes.  

The ZEUS Collaboration announced a preliminary result on the asymmetry measured with their Leading Proton Spectrometer: $A_{LT} = -0.049 \pm 0.058 ({\rm stat}) ^{+0.056}_{-0.009} ({\rm syst})$, with $\tfrac{{\rm d}\sigma^D}{{\rm d}\Delta\phi} \propto 1 + A_{LT} \cos \Delta\phi$~\cite{bib:zeus_lps}.

Since the precision on the electron $\phi$ obtained from SPACAL is extremely good, the resolution on $\Delta\phi$ is entirely determined by the VFPS.  The expectation is that a $\Delta\phi$ distribution with 15 bins can be obtained with 10,000 events in each bin.  This would also allow to measure the asymmetry as a function of $\beta$ or $Q^2$.  Strong variations as a function of these variables are expected, as the higher twist longitudinal photon induced cross section is predicted to dominate at large $\beta$ and small $Q^2$.

\subsubsection*{Deeply Virtual Compton Scattering}

The study of Deeply Virtual Compton Scattering (DVCS) process, i.e.\@ the hard diffractive scattering of a virtual photon off the proton ($ep \rightarrow ep\gamma$) provides access to the skewed parton densities, which are generalizations of the familiar parton distributions of deep-inelastic scattering, but include parton momentum correlations.

An analysis of DVCS based on data collected by H1 in 1997 yielded only 25 events with $Q^2 > 8 {\rm\ GeV}^2$ (out of a total of 100 events with $Q^2 > 2 {\rm\ GeV}^2$)\footnote{With the \mbox{HERA-II} upgrade the inner SPACAL has been removed so that the acceptance is now limited to $Q^2 > 8 {\rm\ GeV}^2$.}.   The expectation for the \mbox{HERA-II} running period is to collect 3600 DVCS events in the same $Q^2$ range.  The VFPS extends the accessible kinematic range toward low $W$ and low $Q^2$, doubling the number of events that can be triggered by the central H1 detector alone.

\section*{Conclusion}

It has been shown that the VFPS is mandatory to effectively trigger on diffractive events during the \mbox{HERA-II} running period.  Without this device it would not be possible anymore to study inclusive diffractive scattering, but also in the semi-inclusive and exclusive channels, the VFPS allows to enlarge the accessible kinematic range and to keep trigger condition as unbiased as possible.

The VFPS will have a very good acceptance in a limited window around $\xpom = 0.01$, while the resolution  on the reconstructed proton momentum is sufficient to allow exciting physics analyses.

The VFPS will be inserted in the HERA machine in early 2003 and will collect data until 2006.
	
\bibliographystyle{unsrt}
\bibliography{proc}
 
\end{document}